\documentclass[%
reprint,
superscriptaddress,
preprintnumbers,
amsmath,
amssymb,
aps,
floatfix,
]{revtex4-2}
\usepackage{graphicx}
\usepackage{dcolumn}
\usepackage{bm}
\usepackage[utf8]{inputenc}
\usepackage[T1]{fontenc}
\usepackage{mathptmx}
\usepackage{textcomp}
\usepackage{siunitx}
\usepackage{bigints}

\usepackage{hyperref}
\usepackage[switch]{lineno}
\usepackage[table,xcdraw,dvipsnames]{xcolor}
\usepackage{booktabs}
\usepackage{float}
\usepackage{color}
\usepackage{ulem}
\usepackage{multirow}

\begin{document}
\title[Reconfigurable three dimensional magnetic nanoarchitectures]{Reconfigurable three dimensional magnetic nanoarchitectures}

\author{Sabri~Koraltan}
\email{sabri.koraltan@tuwien.ac.at}
\affiliation{Physics of 3D Nanomaterials, Institute of Applied Physics, TU Wien, A-1040 Vienna, Austria}
\affiliation{Physics of Functional Materials, Faculty of Physics, University of Vienna, A-1090 Vienna, Austria}

\author{Fabrizio~Porrati}\email{porrati@physik.uni-frankfurt.de}
\affiliation{Goethe University Frankfurt, Frankfurt am Main, Germany}

\author{Robert~Kraft}
\affiliation{Physics of Functional Materials, Faculty of Physics, University of Vienna, A-1090 Vienna, Austria}%
\author{Sven~Barth}
\affiliation{Goethe University Frankfurt, Frankfurt am Main, Germany}

\author{Markus~Weigand}
\affiliation{Institut für Nanospektroskopie,Helmholtz-Zentrum Berlin für Materialien und Energie GmbH,
12489 Berlin, Germany}

\author{Claas~Abert}
\affiliation{Physics of Functional Materials, Faculty of Physics, University of Vienna, A-1090 Vienna, Austria}%
\affiliation{Research Platform MMM Mathematics-Magnetism-Materials, University of Vienna, A-1090 Vienna, Austria.}%

\author{Dieter~Suess}%
\affiliation{Physics of Functional Materials, Faculty of Physics, University of Vienna, A-1090 Vienna, Austria}%
\affiliation{Research Platform MMM Mathematics-Magnetism-Materials, University of Vienna, A-1090 Vienna, Austria.}%

\author{Michael~Huth}
\affiliation{Goethe University Frankfurt, Frankfurt am Main, Germany}

\author{Sebastian~Wintz}\email{sebastian.wintz@helmholtz-berlin.de}%
\affiliation{Institut für Nanospektroskopie,Helmholtz-Zentrum Berlin für Materialien und Energie GmbH,
12489 Berlin, Germany}

\date{\today}

\begin{abstract}

Three-dimensional (3D) nanomagnetism is a rapidly developing field within magnetic materials research, where exploiting the third dimension unlocks opportunities for innovative applications in areas such as sensing, data storage, and neuromorphic computing. Among various fabrication techniques, focused electron beam-induced deposition (FEBID) offers high flexibility in creating complex 3D nanostructures with sub-100 nm resolution. A key challenge in the development of 3D nanomagnets is the ability to locally control the magnetic configuration, which is essential to achieve desired functionalities. In this work, the magnetization reversal mechanism of a three-dimensional nanoarchitecture fabricated using focused electron beam-induced deposition is investigated by combining direct observation via scanning transmission X-ray microscopy with finite element micromagnetic simulations. In particular, our investigation shows that the magnetization of the components of a three-dimensional $\rm Co_3Fe$ tetrapod can be reversed individually and sequentially. Finally, it is demonstrated that complete control and reconfigurability of the system can be achieved by tuning the direction of the applied magnetic field.  

\end{abstract}
\maketitle

\section{\label{sec:intro}Introduction}
Magnetic materials are used in a variety of applications, including hard-disk storage devices~\cite{piramanayagam2009recording}, magnetic field sensors~\cite{suess2018topologically, leitao2024enhanced}, logic circuits~\cite{currivan2016logic}, and unconventional~\cite{finocchio2021promise, finocchio2024roadmap, grollier2020neuromorphic} computing tasks, among others. Predominantly, the design and conceptual frameworks are based on planar thin films, which can be manufactured using physical vapor deposition techniques such as magnetron sputtering~\cite{kelly2000magnetron}, electron beam evaporation~\cite{singh2005review}, pulsed laser deposition~\cite{shepelin2023practical} or molecular beam epitaxy~\cite{joyce1985molecular}. These deposited magnetic specimens are amenable to further processing into functional spintronic devices via micro- and nanofabrication techniques, including photolithography and electron beam lithography. 
In particular, recent advances have drawn significant attention towards three-dimensional magnetism, demonstrating new degrees of freedom available for exploration~\cite{fernandez2017three, makarov2022new,raftrey2022road, fischer2020launching}. Initial works suggest that racetrack memories~\cite{parkin2008magnetic}, high-density storage solutions~\cite{koraltan2021tension}, and magnonic networks~\cite{cheenikundil2022high, sahoo2021observation} have the potential to substantially enhance the energy efficiency and overall performance of future generations of 3D spintronic devices~\cite{gubbiotti20242025}.

Currently, two primary techniques have been emphasized in the experimental realization of these intricate 3D geometrical structures. The first of these methodologies is two-photon lithography (TPL)~\cite{harinarayana2021two}, which, when combined with physical vapor deposition~\cite{donnelly2015element, may2021magnetic, pip2022x} or atomic layer deposition~\cite{guo2023realization} , has been shown to be an efficient and scalable approach to generate complex 3D magnetic lattices and structures. The second promising method is focused electron beam-induced deposition (FEBID) as a direct-write technique~\cite{utke2008gas, huth2012focused, huth2018focused, fernandez2020writing}, in which a focused electron beam interacts with the molecules of a precursor gas within the chamber of a scanning electron microscope~\cite{barth2020precursors}, and the resultant secondary electrons facilitate the direct deposition of the magnetic material at the targeted location. Using this technique, intricately complex magnetic structures have been fabricated, enabling the investigation of phenomena such as 3D artificial lattices~\cite{keller2018direct}, geometrical chirality~\cite{sanz2020artificial}, domain-wall automotion~\cite{skoric2022domain}, complex stray-field textures~\cite{donnelly2022complex}, partial skyrmions~\cite{fullerton2025fractional}, and Bloch points~\cite{fernandez2013three}. It should be noted that FEBID can be used to fabricate scaffolds to be covered by magnetic materials by sputtering~\cite{skoric2022domain} or chemical vapor deposition~\cite{porrati2023site}. Overall, the effective implementation of 3D magnetic functional devices critically hinges on the
ability to precisely control their magnetization configurations. In this regard, the key challenge lies in realizing the reconfigurability of magnetic states within complex 3D nanoconduits, a
capability that is essential to unlock advanced device functionalities~\cite{fullerton2025design, schroder2025origin, gubbiotti20242025}. 

\begin{figure*}[]
    \centering
    \includegraphics[width=\textwidth]{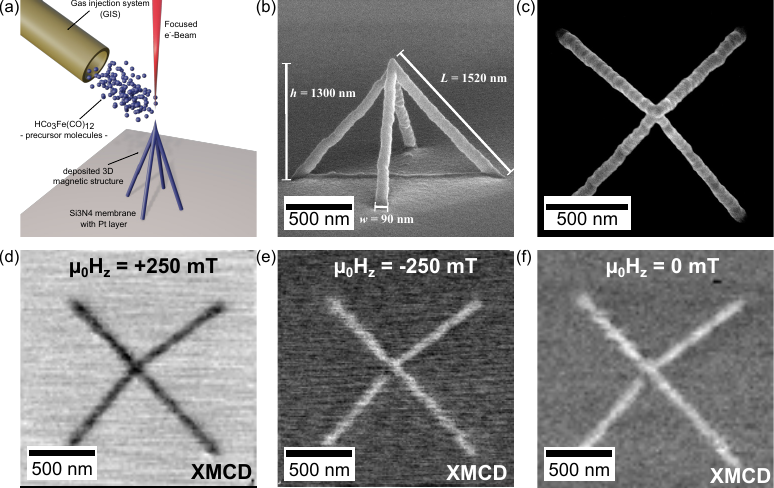}
    \caption{Three dimensional Co$_{3}$Fe nanoarchitecture fabricated using FEBID: (a) Summary the deposition process in which precursor molecules are introduced into the system via a gas injection system (GIS), enabling the formation of a cobalt-iron (Co$_{3}$Fe)-based material under the influence of a focused electron beam. Scanning electron micrographs provided in (b) and (c) present the top and 3D views, respectively. The X-ray magnetic circular dichroism (XMCD) contrast, as assessed using (STXM), is depicted for magnetization orientations saturated with -250 mT (d) and saturated with +250 mT (e) in top view. The remament state is shown in (f).}
    \label{fig:fig01}
\end{figure*}

In this work, we study the magnetic reconfigurability of a three-dimensional nanoarchitecture fabricated directly through FEBID. To investigate the magnetization states within a magnetic tetrapod structure, we utilize scanning transmission X-ray microscopy (STXM), which enables us to discern the individual reversal of each constituent element of the nanoarchitecture. A tetrapod is a rotational symmetric structure with four legs that merge in a single apex; see Fig.~\ref{fig:fig01}. Additionally, we provide experimental evidence demonstrating that the order of magnetic reversal can be modified by adjusting the angle of the applied magnetic field, thereby permitting complete control and reconfigurability. Our results are corroborated by finite-element micromagnetic simulations, through which we also uncover intricate magnetization states embedded within the three-dimensional texture. Our findings mark an important step in the advancement of 3D magnetic functional devices, with prospective applications in 3D magnetic nanoconduits, magnonic crystals, data storage, memory or computing systems~\cite{gubbiotti20242025}.

\section{\label{sec:results}Results}

\subsection{Imaging of saturated and remanent states}
In this work, our objective is to explore the reconfigurability of a three-dimensional nanoconduct with a FEBID-fabricated tetrapod nanostructure. For this purpose, we deposited a tetrapod structure using FEBID. The deposition process and the geometry considered are summarized in Fig.~\ref{fig:fig01}(a). The deposited tetrapod has the height $h = \SI{1.3}{\mu{m}}$, while each leg has the length $L = \SI{1.52}{\mu{m}}$, which was calculated from the 2D projection, which was measured as $L_{\rm{2D}} = \SI{0.8}{{\mu}m}$. The width of each leg is about $\SI{90}{nm}$. SEM images of the tetrapod are shown in Fig.~\ref{fig:fig01}(b) (top view) and Fig.~\ref{fig:fig01}(c) (3D-view), respectively. The SEM images reveal that the legs meet at the apex, which shows a slight increase in thickness. Furthermore, small fabrication defects are visible, and the fabricated legs are not perfectly smooth cylindrical nanowires.

Due to the aspect ratio of each of the legs of the tetrapods, the shape anisotropy will dominate the energetics of the system overall. Thus, we expect the magnetization of the tetrapod to follow the long axis of each leg, potentially forming nonuniform textures at the apex. The magnetic state is probed using scanning transmission X-ray microscopy (STXM) at the MAXYMUS endstation \cite{weigand2022} at BESSY II, Berlin, Germany. The sample was first mounted at normal incidence. That is, the applied magnetic field and the X-ray beam are along the z direction. We first apply a strong magnetic field out-of-plane (oop), and acquire images with both X-ray polarizations. Dividing the two images allows us to obtain the X-ray magnetic circular dichroic (XMCD) contrast. Here, XMCD contrasts give us a measure for the $z$ component of the magnetization. The XMCD data are shown in Fig.~\ref{fig:fig01}(d) and Fig.~\ref{fig:fig01}(e) for positive ($\mu_0H_z = \SI{+250}{mT}$) and negative ($\mu_0H_z = \SI{-250}{mT}$) saturations, respectively. When returning to the remanent state, we observe that the magnetization remains uniform and no switchings are observed, as shown in Fig.~\ref{fig:fig01}f. All STXM images presented in this work were acquired at an X-ray energy corresponding to the onset of the $L_3$ Fe edge at nominally ~ \SI{707.6}{eV}.

\subsection{Succesive magnetization reversal}
The tetrapod is expected to behave magnetically similar to the vertices in 2D and 3D artificial spin ices~\cite{skjaervo2020advances, koraltan2021tension, may2021magnetic}. Thus, the magnetization of each leg can be manipulated by magnetic fields, where the order of switching might be dictated by the nucleation and annihilation of localized magnetic charges~\cite{castelnovo2008magnetic}. To experimentally test these expectations, we acquire XMCD images under different magnetic fields. 

\begin{figure*}[]
    \centering
    \includegraphics[width=\textwidth]{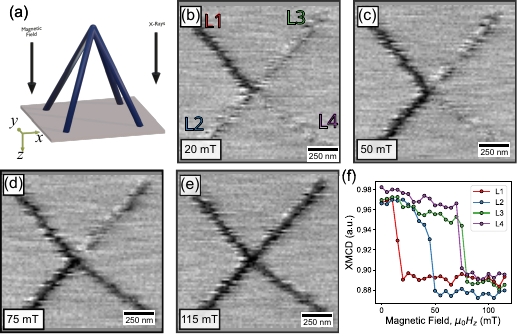}
    \caption{Magnetic characterization of a reconfigurable three-dimensional nanoarchitecture: (a) Depiction of the experimental geometry in which both the magnetic fields and the X-rays are aligned parallel to the $z$-axis.  The magnetic configuration, captured using STXM with XMCD as the contrast formation mechanism, is presented in panels (b-e) (contrast with respect to the saturated state), illustrating individual switching phenomena in response to the applied field. A black contrast indicates downward magnetization, whereas a brighter contrast signifies upward magnetization. (f) The XMCD contrast, averaged locally over the four legs within the nanoarchitecture as a function of the applied magnetic field is shown in (b) and reveals distinct variations in coercivity.}
    \label{fig:fig02}
\end{figure*}

We remain in the same configuration, where both X-rays and magnetic fields are along the $z$ axis; see Fig.~\ref{fig:fig02}(a). The magnetic state is first saturated by applying a strong field $\mu_0H_z = \SI{-250}{mT}$. Then, from the remanance we increase the magnetic field by \SI{5}{mT}. We acquire a STXM image at each field and calculate the XMCD contrast with respect to the saturated state. Selected magnetization states during the field sweep are shown in Fig.~\ref{fig:fig02}, where we show the magnetization configuration at $\SI{20}{mT}$
(Fig.~\ref{fig:fig02}(b)), $\SI{50}{mT}$
(Fig.~\ref{fig:fig02}(c)), $\SI{75}{mT}$
(Fig.~\ref{fig:fig02}(d)), and $\SI{115}{mT}$
(Fig.~\ref{fig:fig02}(e)). The STXM data reveals that the the legs' magnetization are reversed in a succesive manner towards opposing saturation (the final contrast is black). 
From these STXM data, we can obtain locally averaged quantities by masking the data in the legs individually. Figure~\ref{fig:fig02}(f)  shows the averaged XMCD contrast as a function of the applied magnetic field for each leg of the tetrapod. The experimental data clearly show abrupt switches that indicate that magnetization is reversed for strong enough antiparallel fields. Hence, our results indicate that the magnetization in the three-dimensional tetrapod can be controlled by the magnitude of the applied field. In Supplementary Video \#1 the animated step-like sequential switching process as a function of applied field can be observed more closely.

Due to the symmetry of the applied field and the easy axis of each of the elements, the order in which the reversal occurs might be expected to be rather stochastic. However, repeated experiments on this and on a similar sample (Supplementary Video \#2) demonstrate that the control of the reversal, as well as the order of switching of the different legs, was reproducible. A possible explanation is that each leg might contain structural or chemical defects such as local variation of metal content, thus locally altering the coercive fields of each element. The differences in the coercivities become rather a feature, because it allows to \textit{selectively} reverse the magnetization by tuning the magnitude of the magnetic field. The reproducibility of the results suggests that the magnetization of a 3D nanostructure can be controlled by the magnitude of the magnetic field. 

A second possible and perhaps complementary scenario involves the presence of topologically non-trivial spin textures and magnetic solitons, which govern and dictate the reversal mechanisms. Volkov et al.~\cite{volkov2024three} have shown that a tetrapod structure - topologically equivalent to the one investigated in this work - is homeomorphic to a
sphere. Thus, the total vorticity of the sample is expected to be $Q^{\Sigma} = +2$ . This could be achieved, for example, by four vortices and two antivortices, which might appear as surface textures. In this case, the presence and location of topological textures might influence the reversal mechanisms, where reversing the magnetization of one leg might involve annihilation and nucleation of non-trivial textures.

\begin{figure*}[]
    \centering
    \includegraphics[width=0.9\textwidth]{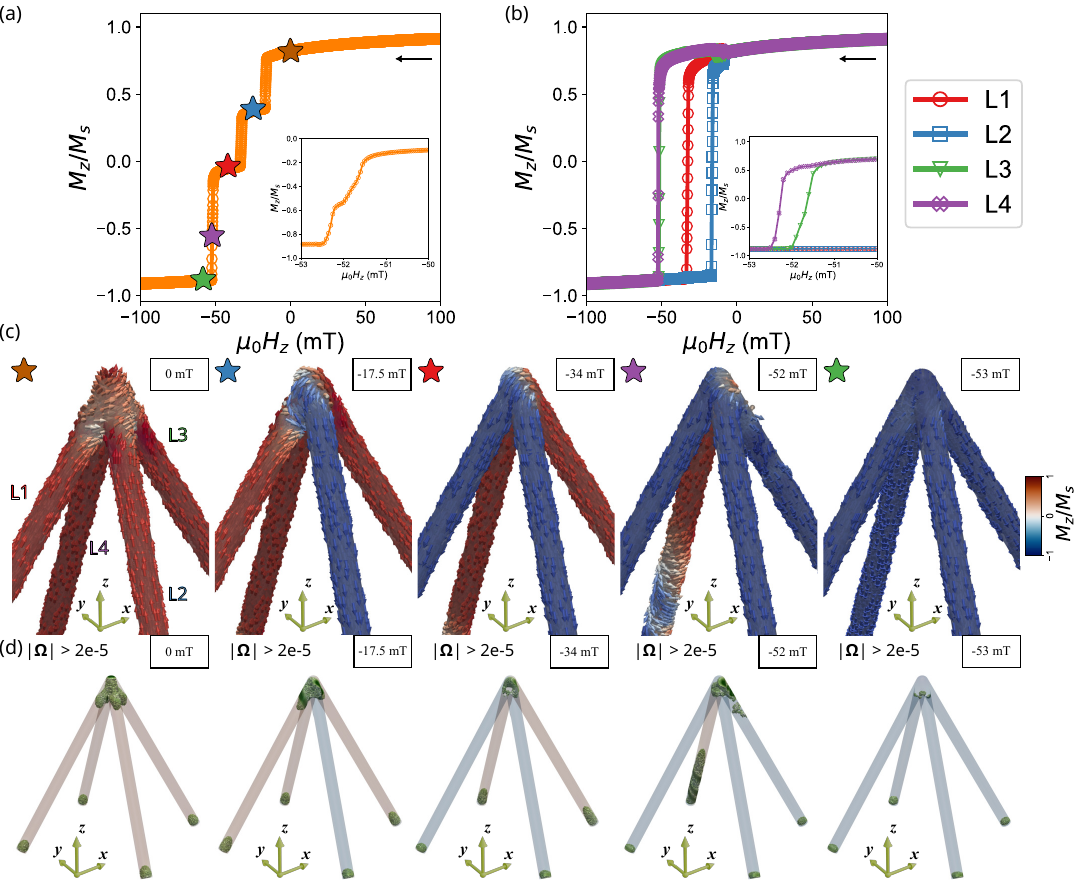}
    \caption{The simulated hysteresis loop depicted in (a) shows the total $M_z$ component of the magnetization (normalized), averaged over the entire nanoarchitecture, and plotted against the applied magnetic field. The black arrow shows the direction in which the magnetic field is swept. The inset figure shows a zoomed-in, which highlights that the third and fourt switch happend at similar fields, yet are separated events. In panel (b), the individual contributions from each leg are presented. Panel (c) shows the magnetic configurations at specific fields, which are indicated by stars in (a), highlighting the successive individual magnetization switchings. The colors (red, blue) indicate the $M_z$ component of the magnetization. Panel (f) illustrates $|\Omega|$ thresholded for values $|\Omega| > \SI{2e-5}{}$, that highlight the presence of topological texture formations at ends of the legs and at the apex. All states are transient states as the magnetic field is time-dependent.}
    \label{fig:fig03}
\end{figure*}

To gain a deeper understanding of the magnetization states and reversal mechanism, we performed micromagnetic simulations using the finite element simulation package magnum.pi~\cite{abert2013magnum}. For this, we create a 3D CAD geometry and mesh it using tetrahedral elements with \SI{6}{nm} mesh length; see Sect.~\ref{sec:methods} for more details.
First, we simulate the response of the magnetic state to an out-of-plane (oop) external field along $\mathbf{z}$. The simulated M-H curve for the entire structure is shown in Fig.~\ref{fig:fig03}(a). In this representation, a clear staircase-like hysteresis is observed, where three major switching events are observed. This strongly suggests the presence of individual reversal events in the system. A closer look at the third switching event (inset of Fig.~\ref{fig:fig03}(a)) reveals, in fact, two separate events that are very close in field. When we plot the averaged magnetization for each element individually (Fig.~\ref{fig:fig03}(b)), indeed we observe that the magnetization reverses successively for all legs, which is in excellent agreement with our experimental findings. Moreover, the final two reversal fields are rather similar in magnitude, an aspect that we have also observed experimentally. 

The simulated magnetic states are given for selected fields (colored stars in (a)) in Fig.~\ref{fig:fig03}(c), where the arrows show the normalized magnetization direction colored according to the $M_z$ component of the magnetization. At remanence, all legs are positively magnetized, where a non-uniform topological vortex-like texture is visible at the apex. To quantify the topology of the magnetization, we calculate the norm of the topological charge flux density, given as

\begin{equation}
    \boldsymbol{\Omega}_\alpha = \dfrac{1}{8\pi}\epsilon_{\alpha\beta\gamma}\epsilon_{ijk}m_i{\partial}_\beta{m_j}{\partial}_\gamma{m_k},
\end{equation}
where $m_i$ denotes the components of the normalized magnetization vector.  When integrated over the closed 2D surface of the magnetic structure, the topological winding number (skyrmion number) \cite{finocchio2016magnetic} is obtained. The topological charge flux density $\boldsymbol{\Omega}$ is thus a measure of local topological winding, indicating the presence of vortex- and antivortex-like structures. Figure~\ref{fig:fig03}(d) shows the norm of $|\boldsymbol{\Omega}|$, which shows the presence of nonuniform-textures at the ends and at the apex at remanence. In Supplementary Fig. S1, we show that magnetic vortices are indeed stable at the end of the legs, while more complex surface textures are forming on and around the apex; see Supplementary Fig.~S2. The bulk texture of the apex also includes a magnetic vortex, which can be visualized by taking 2D slices in the $xy$-plane, as done in Supplementary Fig.~S3. The analysis of possible surface antivortices and the quantification of the total vorticity of the nanostructure is rather complex and is beyond the scope of this work. 

\begin{figure*}[]
    \centering
    \includegraphics[width=\textwidth]{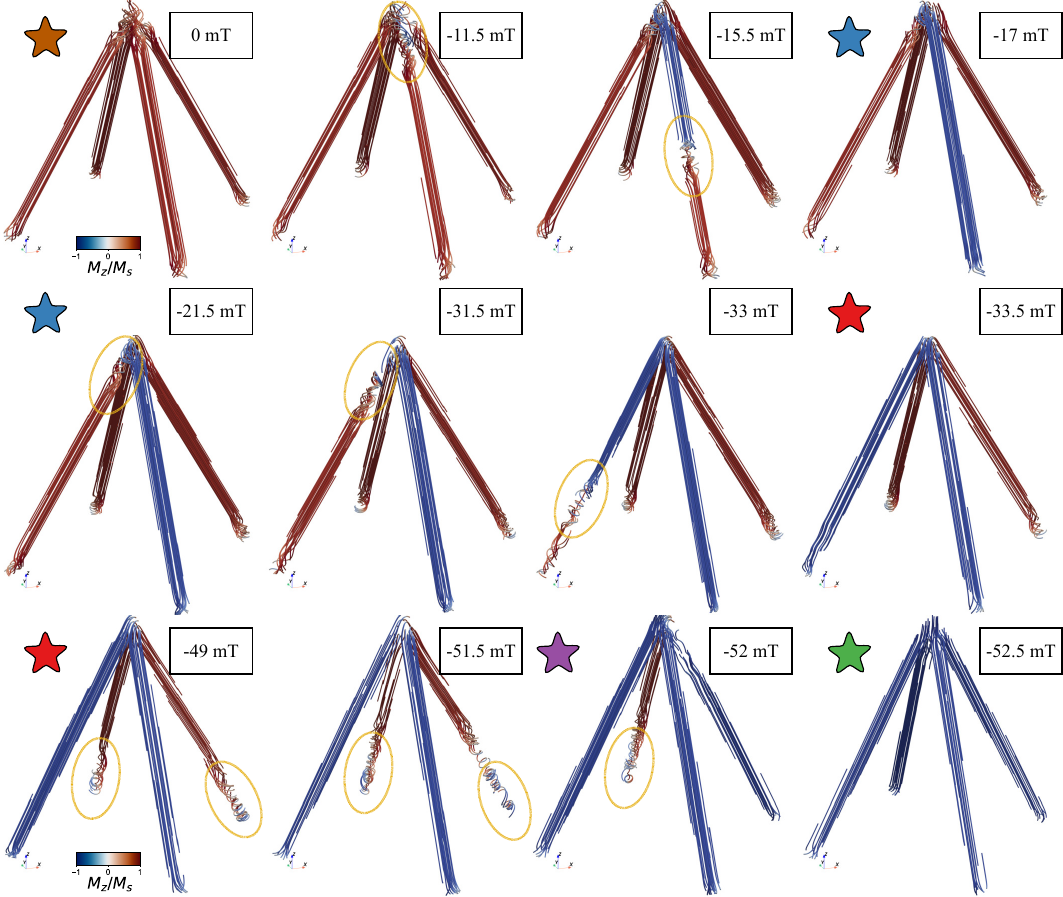}
    \caption{Magnetization flow at different external fields $\mu_{0}H_z$ which shows the reversal mechanisms dominated by vortex wall propagations and nucleations. The stream tracer used here highlights the localized nonuniformity and winding of the magnetization. The dashed yellow ellipses highlight the location of magnetic vortices, whose presence dictates which leg will reverse next. The colored stars refer to Fig.~\ref{fig:fig03}.}
    \label{fig:fig04}
\end{figure*}

When the applied field is further decreased towards negative saturation, the legs of the tetrapod reverse individually. 
Further comparing the experiments with the simulations, we observe that after the reversal of the first leg, the second reversing leg is never opposing the initial one but is adjacent to it. As in any artificial spin system, the order of reversal could be dictated by the creation of magnetic charges at the apex~\cite{skjaervo2020advances, koraltan2021tension}. Thus, in our system, the non-uniform topological magnetic textures might also play a key role. When comparing the magnetization states in Fig.~\ref{fig:fig03}(c) and the localizations of the magnetic nonuniformities in Fig.~\ref{fig:fig04}(d), we recognize a correlation between the localization of the solitons and the order of reversal, especially of the first two legs. To further analyze this hypothesis, we plot the magnetization in Fig.~\ref{fig:fig04} using streamlines, which is an instructive way to visualize the local winding of the magnetization. The field-dependent evolution of the magnetization reveals that for the first reversal, a magnetic vortex is nucleated at the apex ($\mu_0H_z = \SI{-11.5}{mT}$), which then propagates to the end ($\mu_0H_z = \SI{-15.5}{mT}$), subsequently reversing leg L2 of the tetrapod ($\mu_0H_z = \SI{-17}{mT}$). The second reversal is then fully governed by the imbalance of magnetic charges at the apex, which results in a vortex wall slightly shifted towards the other leg L1 ($\mu_0H_z = \SI{-21.5}{mT}$). The vortex wall first fully shifts to L1 ($\mu_0H_z = \SI{-31.5}{mT}$), and then propagates ($\mu_0H_z = \SI{-33}{mT}$) through the leg to fully reverse the magnetization ($\mu_0H_z = \SI{-33.5}{mT}$). In the case of the final two reversals, we observe that there is no vortex wall left at the apex. Thus, it appears that the magnetization reversal is governed by the nucleation of vortex walls at the ends of the legs L2 and L3, which now propagate up towards the apex. Our numerical investigation suggests that, based on the magnetic configuration of the apex, there are two main reversal mechanisms: propagation of a domain wall injected from the apex and nucleation and propagation of a domain wall from the bottom point of a leg. Overall, our numerical simulations suggest that the local magnetic topology dictates the reversal mechanisms and order of reversal in our 3D magnetic nanoarchitecture.

\subsection{Angle dependent selective reversals}

\begin{figure*}[]
	\centering
	\includegraphics[width=\textwidth]{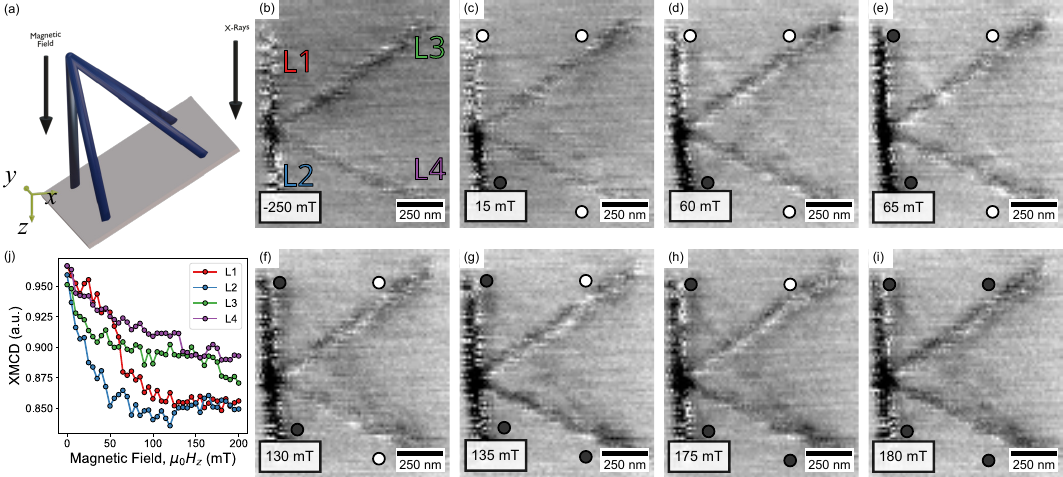}
	\caption{The experimental magnetization reversal for a specific angle of applied magnetic field: (a) The experimental configuration is illustrated, in which the sample was rotated $30\, ^\circ$ around the x-axis, thus altering the magnetic field component parallel to the legs. The XMCD images for varying magnetic fields are presented in (b-i) (contrast with respect to the saturated state), illustrating the switching of the magnetization in all legs. The legs that are more parallel to the applied field switch sooner. The contrast of the other two legs (L3, L4) decreases as we are less sensitivie to the magnetization component parallel to those legs. Succesive magnetic fields are shown to highligh better the states before (d,f,h) and after (c,e,g,i) the reverals. The XMCD contrast, averaged individually across the four elements within the nanoarchitecture as a function of the applied magnetic field, is depicted in (j). The circles in (b-i) indicate wether the legs have switched (black) or not (white). We refer to Supplementary Video \#3 for a clear recognition of the succesive switchings.}
	\label{fig:fig05}
\end{figure*}

After demonstrating experimentally and by simulations that the magnetization of the legs can be reversed individually by application of magnetic fields, our goal is to achieve selective switching of particular legs, enabling full control and reconfigurability of the 3D magnetization of the tetrapod. For this purpose, we change the measurement geometry. The sample is now mounted such that there is a $\Theta = \SI{30}{^\circ}$ angle between the X-rays and the samples' normal orientation; see Fig.~\ref{fig:fig05}(a). Thus, the corresponding image is now the projection of the sample along $\left( \sin(\SI{30}{^\circ}),\, 0,\, \cos(\SI{30}{^\circ})\right)$. In Figs. ~\ref{fig:fig05}(b-i), we show the measured XMCD images at different magnetic fields, which support the claim that the magnetization of the elements more parallel to the field can be reversed at lower fields. Considering the reversals of the legs displayed in Figs. ~\ref{fig:fig05}(f-i), we need to remark that, because of the tilt, we are now less sensitive to the magnetization component, which is parallel to the legs. Thus, a considerable loss of contrast is visible, which makes it challenging to visually recognize the successful switching events. The black and white circles in the images are a guide to the eye for which element a switching is observed. Comparing subsequent images before and after the reversal allows a better distinction if the legs have switched or not. In addition, the animation provided in the Supplementary Video~\#3 clearly shows that all four legs switch in a successful manner, while the legs parallel to the field reverse first.

The quantified (averaged) local XMCD signal as a function of the total magnetic field is shown in Fig.~\ref{fig:fig05}(f). This time, it appears that elements (L1,L2) are reversing their magnetization again with a rather abrupt switching event, while the other two elements (L3,L4) are experiencing a delayed reversal process. For L3 and L4 the overall drop in the signal is not very clear due to the loss of sensitivity.

\begin{figure*}[]
    \centering
    \includegraphics[width=\textwidth]{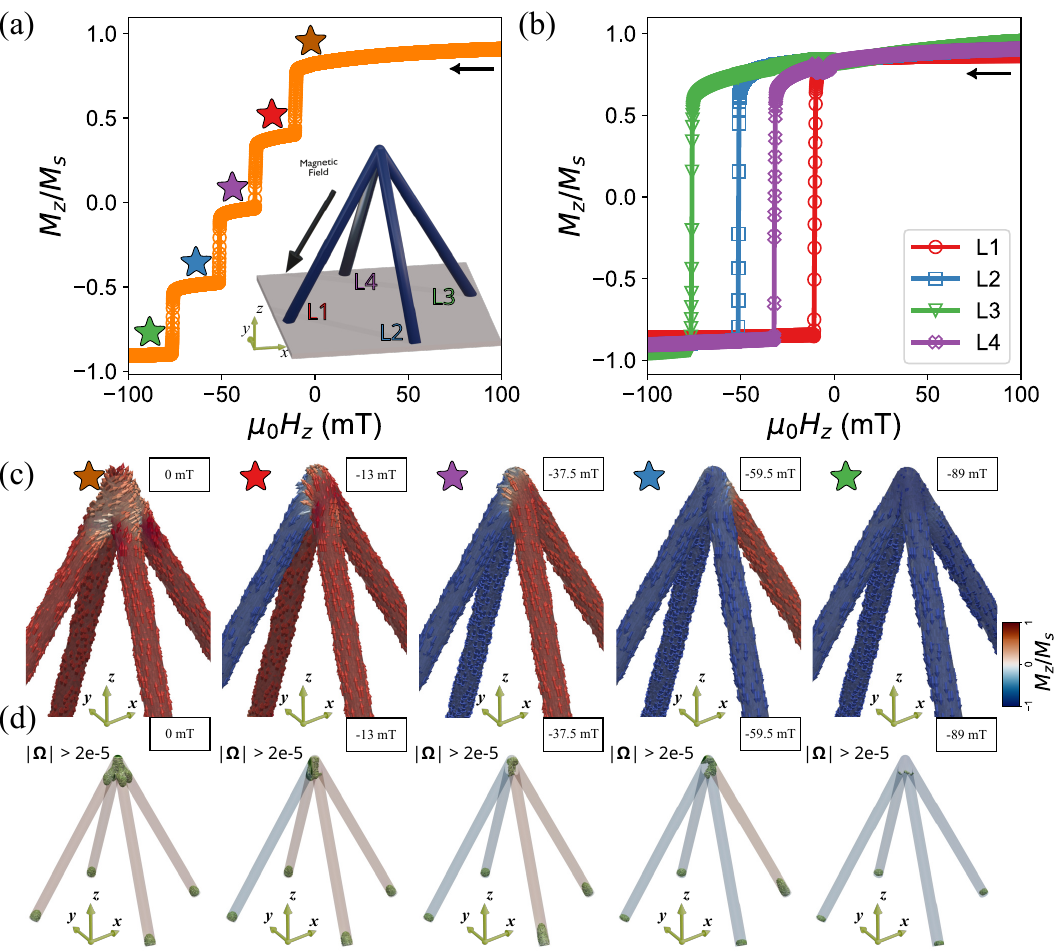}
    \caption{Simulated hysteresis loops for magnetic field applied along $\mathbf{H} = (\sin{\theta}\cos{\phi}~, \sin{\theta}\sin{\phi}~, \cos{\theta})$, with $\theta = 30^\circ$, $\phi = 0^\circ$, where in (a) the magnetization is averaged over the entire nanoarchitecture and in (b) is resolved for each leg individually. The magnetic configurations are shown in (c) for the four steps in the hysteresis loops shown in (a), where the element parallel to the field switches its magnetization first. The norm $|\boldsymbol{\Omega}|$ in (d) shows once more the presence of the topologically nontrivial magnetization states.} 
    \label{fig:fig06}
\end{figure*}

The experimental setup used for this study limited the geometries that can be investigated to 0 or 30 degrees with respect to the sample normal at a fixed azimuthal angle. Thus, to test the full reconfigurability of the tetrapod, we turn again to micromagnetic simulations. The results are summarized in Fig.~\ref{fig:fig06}. First, the magnetic field is applied along $\theta = 30^\circ$ and $\phi=0^\circ$, which is parallel to the leg L1, as shown in the inset of Fig.~\ref{fig:fig06}(a), where we also illustrate the total averaged hysteresis M-H loop. In contrast to the previous case from Fig.~\ref{fig:fig03}, the hysteresis loop now clearly shows four switching events. From this, we understand that the switching fields of all legs are tunable. The magnetization averaged over the individual legs shown in Fig.~\ref{fig:fig06}(b), accompanied by the magnetic configurations in Fig.~\ref{fig:fig06}(c) and $|\boldsymbol{\Omega}|$ in Fig.~\ref{fig:fig06}(d) shows that the L1's magnetization reverses first. The switching fields in this geometry are more distinct, allowing for the stepwise reversal to be clearly visible in the total hysteresis loop. A closer look at the magnetization reversal confirms that, also for this case, the locations of vortex walls dictate the order of reversals.

For the sake of completeness, we also show in the Supplementary Fig. S4 that the magnetization of arbitrary legs can also be controlled. For instance, we apply the magnetic field parallel to L2, thus, $\theta = 30^\circ$ and $\phi=90^\circ$. Supplementary Figure~S4 is analogous to Fig.~\ref{fig:fig06}, and demonstrates the angle-dependent selective and succesive magnetization reversals in the tetrapod.

Furthermore, we repeat our simulations for multiple combinations of angles $(\theta, \phi)$ and quantify the reversal fields of the simulated hysteresis loops for each leg individually, as shown in the Supplementary Fig.S5. Our simulations indicate that, for each of the legs, an angle-dependent reduction of the coercivity occurs at favored angles. The reversals for unfavored angles (or favored angles for the other legs) seem to occur rather stochastically.

\section{\label{sec:conclusion} Conclusion}
In summary, we investigated the magnetization reversal mechanisms in a three-dimensional nanoarchitecture fabricated through focused electron beam-induced deposition, utilizing direct observation via scanning transmission X-ray microscopy and supported by finite element micromagnetic simulations. Our experimental findings revealed that the magnetization of the legs of a three-dimensional tetrapod can be reversed individually in a sequential manner. Here, the presence and location of nonuniform topological magnetic textures are dictating the order of reversal. Moreover, we demonstrated that complete control and reconfigurability of the system can be obtained by altering the angle of the applied magnetic field. Our results demonstrate experimentally a local control of the magnetization in a three-dimensional architecture. These findings have significant implications for a broad range of potential applications of three-dimensional nanomagnets, including three-dimensional control of magnetism for storage devices, logical operations, sensing, and actuator technologies.

\section{\label{sec:methods} Experimental Section}
\textit{Sample Fabrication.} For being able to perform soft X-ray transmission experiments on the fabricated structures, the tetrapods were grown on $\SI{200}{nm}$ silicon nitrate membranes. Before deposition of the actual tetrpods, a $\SI{5}{nm}$ Pt layer was deposited on top of the membrane to ensure better electron diffusion and avoid undesired depositions. The tetrapods were fabricated by a dual-beam focused ion beam / scanning electron microscope (Nova Nanolab 600, FEI) equipped with a Schottky electron emitter. In FEBID, the adsorbed molecules of a precursor gas dissociate by interaction with the electron beam. The solid part of the dissosation product generates the sample by the action of the electron beam\cite{huth2018focused}. For the present fabrication, the precursor gas HCo$_3$Fe(CO)$_{12}$ was used~\cite{porrati2015direct, keller2018direct, volkov2024three, schroder2025origin}. The precursor was injected into the SEM using a standard gas injection system through a capillary with an inner diameter of $\SI{0.5}{mm}$. During deposition, the precursor temperature was 64 ° C, the capillary surface distance was about $\SI{100}{\mu m}$ and the capillary incidence angle was $40^\circ$. The base pressure of the SEM was $\SI{4.2e-7}{mbar}$, which increased to approximately $\SI{4.4e-7}{mbar}$ during deposition. The tetrapods were grown in a heuristic approach~\cite{porrati2015direct}. The electron beam parameters were $\SI{4}{keV}$ for the acceleration voltage and $\SI{4}{pA}$ for the electron beam current. The four branches of the nanostructure were written by dwelling the electron beam for $\SI{10}{ms}$ on each branch, before moving on to the next branch. The lateral speed of the electron beam was $\SI{3.7}{nm/s}$.

\textit{Scanning Transmission X-ray Microscopy}: The STXM measurements were performed at the MAXYMUS end station, BESSY II, Berlin. A highly coherent and brilliant soft X-ray is provided by the synchrotron. The X-rays are focused onto the sample by a diffracting Fresnel zone plate. Undiffracted X-rays and higher-order diffractions are stopped by the central stop of the zone plate and an order sorting aperture (OSA). The sample is raster scanned using a piezo stage, while the transmitted X-rays are collected by a point detector, enabling a lateral resolution of approximately 25 nm. The sample was first mounted in normal incidence (0 degrees), thus providing sensitivity to $M_z$. The energy of the X-rays was varied around the Fe L$_3$ edge to achieve the best possible XMCD contrast. As a consequence of the relatively high thickness of the sample, the ideal measurement conditions for Fe were found at the onset of the L$_3$ edge at a nominal photon energy of $\SI{707.6}{eV}$. The photons can be transmitted very easily through the blank membrane, but the energy is absorbed if they interact with the tetrapod. The signal containing information about the topography of the sample can be eliminated by acquiring an image with positive circularly polarized light and an image with negative circular polarization and dividing them by each other, leaving behind only the signal corresponding to the X-ray magnetic circular dichroism (XMCD) contrast. Thus, alternating contrasts are obtained if a projection of the magnetization is either pointing parallel to the propagation direction of the soft X-rays or antiparallel to that.

\textit{Micromagnetic simulations}: The numerical simulations were performed using the finite element simulation software magnum.pi~\cite{abert2013magnum}. The geometry is created and meshed using SALOME. The discretization length was chosen as $\SI{6}{nm}$. The material parameters used are a saturation magnetization of $M_s = \SI{700}{kA/m}$ and and an exchange stiffness constant of $A_{\rm ex} = \SI{12}{pJ/m}$. The Landau-Lifshitz-Gilbert (LLG) equation is solved at a moderate Gilbert damping of $\alpha = 0.1$, which is rather realistic for the CoFe system deposited by FEBID. In the effective field of the LLG, we include the contributions from exchange and Zeeman energy, as well as the demagnetization energies. The demagnetization field is calculated by employing a fast-multiple method, which allows parallelization on multiple processors. The initial magnetization state is chosen to be parallel to the z direction. The external magnetic field is varied between $\mu_0H = \SI{250}{mT}$ and $\mu_0H = \SI{-250}{mT}$ in one microsecond. Magnetization snapshots are saved every nanosecond (every $\SI{0.5}{mT}$). The direction of the magnetic field is varied using the polar and azimuthal angles, $\theta$ and $\phi$, respectively.

\subsection*{Acknowledgements}
The authors thank Kevin Hofhuis for fruitful discussions and the deposition of the platinum layer on the membranes. Furthermore, we thank Sina Mayr, Alexander Kuprava and Oleksii Volkov for fruitful discussions. We thank Simone Finizio for implementing interferometric position correction of the STXM images at the Maxymus beamline. S.K, F.P, S.W. gratefully acknowledge funding through the IEEE Magnetics Society for funding special projects. S.K. acknowledges funding from the European Research Council (ERC) under
the European Union’s Horizon 2020 research and innovation programme, Grant Agreement No. 101001290
(3DNANOMAG). C.A. and R.K acknowledge FWF Vladimir (10.55776/P34671). S.B.: thanks the Deutsche Forschungsgemeinschaft DFG for funding BA6595/5-1. Support through the Frankfurt Center of Electron Microscopy (FCEM) is
gratefully aknowledged. The computational results presented have been achieved using the Vienna Scientific Cluster (VSC). The authors thank Helmholtz-Zentrum Berlin for the allocation of synchrotron radiation beamtime.

\subsection*{Conflict of Interest}
The authors declare no conflict of interest.

\subsection*{Data Availability Statement}
The data that support the findings of this study are available from the corresponding author upon reasonable request.

\subsection*{Keywords}

\bibliography{main}

\end{document}